\documentclass[
    ,final            
  ]
  {aipproc}

\layoutstyle{6x9}

\begin{document}

\newcommand{\lapprox}{%
\mathrel{%
\setbox0=\hbox{$<$}
\raise0.6ex\copy0\kern-\wd0
\lower0.65ex\hbox{$\sim$}
}}
\newcommand{\gapprox}{%
\mathrel{%
\setbox0=\hbox{$>$}
\raise0.6ex\copy0\kern-\wd0
\lower0.65ex\hbox{$\sim$}
}}

\def\nn{\nonumber}
\def\gsim{\ \rlap{\raise 2pt\hbox{$>$}}{\lower 2pt \hbox{$\sim$}}\ }
\def\lsim{\ \rlap{\raise 2pt\hbox{$<$}}{\lower 2pt \hbox{$\sim$}}\ }
\def\dslash{\kern-4pt \not{\hbox{\kern-2pt $\partial$}}}
\def\pslash{\not{\hbox{\kern-2pt p}}}
\def\p{{\bf p}}
\def\q{{\bf q}}
\def\gev{{\rm GeV }}
\def\l{{\rm L}}
\def\db{{$\delta b$ }}
\def\dc{{$\delta c$ }}

\def\evsq{{${\rm eV^2}$ }}
\def\L{{\rm L }}
\def\E{{\rm E }}

\def\pmutau{{${P_{\mu \tau}}$ }}
\def\petau{{${P_{e \tau}}$}}
\def\pmue{{${P_{\mu e}}$ }}
\def\pemu{{${P_{e \mu }}$ }}
\def\pmumu{{${P_{\mu \mu}}$ }}
\def\pee{{${P_{ee}}$ }}

\def\numutonutau{{${\rm {\nu_\mu \to \nu_\tau}}$ }}
\def\numutonue{{${\rm {\nu_\mu \to \nu_e}}$ }}
\def\numutonumu{{${\rm {\nu_\mu \to \nu_\mu}}$ }}

\def\nuetonutau{{${\rm {\nu_e \to \nu_\tau}}$ }}

\def\pnumutonue{{${\rm P ({\nu_\mu \to \nu_e})}$ }}
\def\pnumutonumu{{${\rm P ({\nu_\mu \to \nu_\mu})}$ }}

\newcommand{\nue}{\nu_e}
\newcommand{\be}{\begin{eqnarray}}
\newcommand{\ee}{\end{eqnarray}}
\newcommand{\etal}{{\it et al.}}
\def\anue{{\bar\nu_e}}
\def\numu{{\nu_{\mu}}}
\def\anumu{{\bar\nu_{\mu}}}
\def\nutau{{\nu_{\tau}}}
\def\anutau{{\bar\nu_{\tau}}}
\newcommand{\dm}{\mbox{$\Delta m_{21}^2$~}}
\newcommand{\st}{\mbox{$\sin^{2}2\theta$~}}
\newcommand{\thsol}{\mbox{$\theta_{12}$~}}
\newcommand{\sth}{\mbox{$\sin^{2}2\theta_{13}$~}}
\def\lsim{\:\raisebox{-0.5ex}{$\stackrel{\textstyle<}{\sim}$}\:}
\def\gsim{\:\raisebox{-0.5ex}{$\stackrel{\textstyle>}{\sim}$}\:}
\newcommand{\ms}{\Delta m^2_{21}}
\newcommand{\ma}{\Delta m^2_{31}}
\newcommand{\ts}{\sin^2 2\theta_{\odot}}
\newcommand{\sss}{\sin^2 \theta_{12}}
\newcommand{\sch}{\sin^2 \theta_{13}}
\newcommand{\stch}{\sin^2 2\theta_{13}}
\newcommand{\sa}{\sin^2 \theta_{23}}
\newcommand{\sta}{\sin^22 \theta_{23}}
\newcommand{\mst}{\Delta m^2_{21}{\mbox {(true)}}}
\newcommand{\mat}{\Delta m^2_{31}{\mbox {(true)}}}
\newcommand{\tst}{\sin^2 2\theta_{\odot}{\mbox {(true)}}}
\newcommand{\ssst}{\sin^2 \theta_{12}{\mbox {(true)}}}
\newcommand{\scht}{\sin^2 \theta_{13}{\mbox {(true)}}}
\newcommand{\stcht}{\sin^2 2\theta_{13}{\mbox {(true)}}}
\newcommand{\sat}{\sin^2 \theta_{23}{\mbox {(true)}}}
\newcommand{\stat}{\sin^22 \theta_{23}{\mbox {(true)}}}
\newcommand{\tmt}{$\theta_{23}$}
\newcommand{\tet}{$\theta_{13}$}
\newcommand{\tem}{$\theta_{12}$}
\newcommand{\D}{\Delta}
\newcommand{\pmm}{P_{\mu\mu}}
\newcommand{\sig}{$3\sigma$}
\newcommand{\bb}{$\beta$-beam~}
\newcommand{\bbf}{$\beta$-beam}
\def\ltap{\ \raisebox{-.4ex}{\rlap{$\sim$}} \raisebox{.4ex}{$<$}\ }
\def\gtap{\ \raisebox{-.4ex}{\rlap{$\sim$}} \raisebox{.4ex}{$>$}\ }


\vspace*{-1.8cm} 
\begin{flushright}
HRI-P-07-07-004
\end{flushright}

\vspace*{-0.7cm} 
\title{Magic Baseline Beta Beam}

%

\author{Sanjib Kumar Agarwalla\footnote{S. K. Agarwalla presented this work
at the International Workshop on Theoretical High Energy Physics (IWTHEP 2007), 
Roorkee, India, 15-20 March, 2007.}}{
  address={Harish-Chandra Research Institute, Chhatnag Road, Jhusi, Allahabad
  - 211019, India}
}

\author{Sandhya Choubey}{
   altaddress={Harish-Chandra Research Institute, Chhatnag Road, Jhusi,
  Allahabad - 211019, India}
}

\author{Amitava Raychaudhuri}{
  altaddress={Harish-Chandra Research Institute, Chhatnag Road, Jhusi,
  Allahabad - 211019, India} 
}

\begin{abstract}
We study the physics reach of an experiment where
neutrinos produced in a beta-beam facility at CERN are observed
in a large magnetized iron calorimeter (ICAL) at the India-based
Neutrino Observatory (INO). The CERN-INO distance is close to the
so-called ``magic" baseline which helps evade some of the parameter
degeneracies and allows for a better measurement of the neutrino mass
hierarchy and $\theta_{13}$.

\end{abstract}

\maketitle


To pin down the structure of the neutrino mass matrix, we
need information on the third mixing angle $\theta_{13}$,
the sign$\footnote{The neutrino mass hierarchy is
termed ``normal'' (``inverted'') if $\ma = m_3^2 - m_1^2$ is
positive (negative).}$ of $\Delta m^2_{31}\equiv m_3^2 - m_1^2$ ($sgn(\ma)$)
and the CP phase ($\delta$). The $\nu_e \to \nu_\mu$
transition probability ($P_{e \mu}$) is dependent on all these
three parameters and is termed the ``golden channel'' \cite{golden}
for measuring these unknowns in long baseline
accelerator based experiments. Here we focus on a long baseline
$\beta$-beam \cite{zucc} experiment
in conjunction with a magnetized iron
calorimeter detector with charge identification capability.
The proposal for a detector of this type (ICAL) is being evaluated
by the INO collaboration \cite{ino}. We consider the beta beam source
to be located at CERN. The large baseline captures a matter-induced
contribution to the oscillation probability,
essential for probing $sgn(\ma)$.
The CERN-INO distance happens to be near the so-called
`magic' baseline \cite{magic, eight, magic2} for which the results are relatively
insensitive to the yet unconstrained CP phase. This permits such an
experiment to make precise measurements of the mixing angle $\theta_{13}$
avoiding the issues of intrinsic degeneracy \cite{degeneracy} 
and correlations \cite{golden} which plague other baselines.


The expression for $P_{e \mu}$ in matter,
upto second order in the small parameters $\alpha \equiv
\Delta m_{21}^2/\Delta m_{31}^2$ and $\sin
2\theta_{13}$,  is \cite{golden}:
{\footnotesize{
\begin{eqnarray}
P_{e\mu} & \simeq & \sin^2 2\theta_{13} \, \sin^2 \theta_{23}
\frac{\sin^2[(1- \hat{A}){\Delta}]}{(1-\hat{A})^2}
\pm \alpha  \sin 2\theta_{13} \,  \xi \sin \delta
\sin({\Delta})  \frac{\sin(\hat{A}{\Delta})}{\hat{A}}
\frac{\sin[(1-\hat{A}){\Delta}]}{(1-\hat{A})}
\nonumber \\
&+& \alpha  \sin 2\theta_{13} \,  \xi \cos \delta
\cos({\Delta})  \frac{\sin(\hat{A}{\Delta})}{\hat{A}}
\frac{\sin[(1-\hat{A}){\Delta}]} {(1-\hat{A})}
+ \alpha^2 \, \cos^2 \theta_{23}  \sin^2 2\theta_{12}
\frac{\sin^2(\hat{A}{\Delta})}{\hat{A}^2};
\label{eqn:magic}
\end{eqnarray}
}}
where $\Delta \equiv \Delta m_{31}^2 L/(4 E)$, $\xi \equiv \cos\theta_{13} \,
\sin 2\theta_{12} \, \sin 2\theta_{23}$,  and $\hat{A} \equiv \pm (2 \sqrt{2}
G_F n_e E)/\Delta m_{31}^2$.  $G_F$ and $n_e$ are the Fermi coupling constant
and the electron density in matter, respectively. The sign of the second term
is positive (negative) for $\nu_e \rightarrow \nu_\mu$
($\nu_\mu \rightarrow \nu_e$) channel. The sign of $\hat{A}$ is
positive (negative) for neutrinos (anti-neutrinos) with normal hierarchy
and it is opposite for inverted hierarchy.
When $\sin(\hat{A}\Delta)=0$, the last three terms in Eq. (\ref{eqn:magic}) 
drop out and the $\delta$ dependence disappears from the $P_{e\mu}$ channel, 
which provides a clean ground for $\theta_{13}$ and $sgn(\ma)$ measurement.

Since $\hat{A}\Delta = \pm  (2 \sqrt{2} G_F n_e L)/4$
by definition, the first non-trivial solution for the condition, 
$\sin(\hat{A}\Delta)=0$ reduces to $\rho L = \sqrt{2}\pi/G_F Y_e$,
where $Y_e$ is the electron fraction inside the earth. This
gives $\frac{\rho}{[{\rm g/cc}]}\frac{L}{[km]} \simeq 32725~$,
which for the PREM \cite{prem} density profile of the earth
is satisfied for the ``magic baseline'' \cite{magic, eight, magic2},
$L_{\rm magic} \simeq 7690 ~{\rm km}$.
The CERN-INO distance corresponds to $L=7152$ km,
which is tantalizingly close to this magic baseline.

\begin{figure}[!t]

\includegraphics[height=.24\textheight]{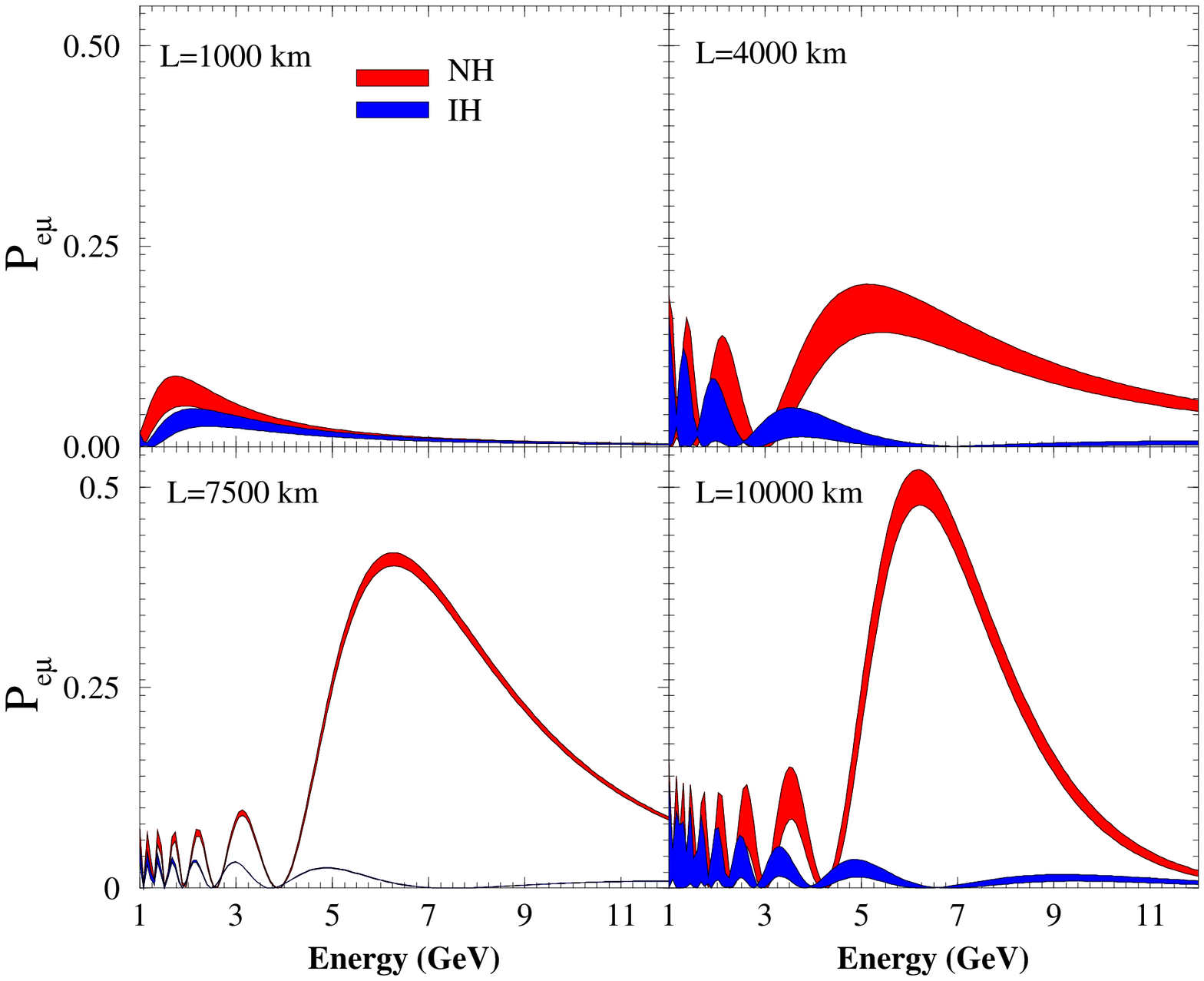}
\includegraphics[height=.24\textheight]{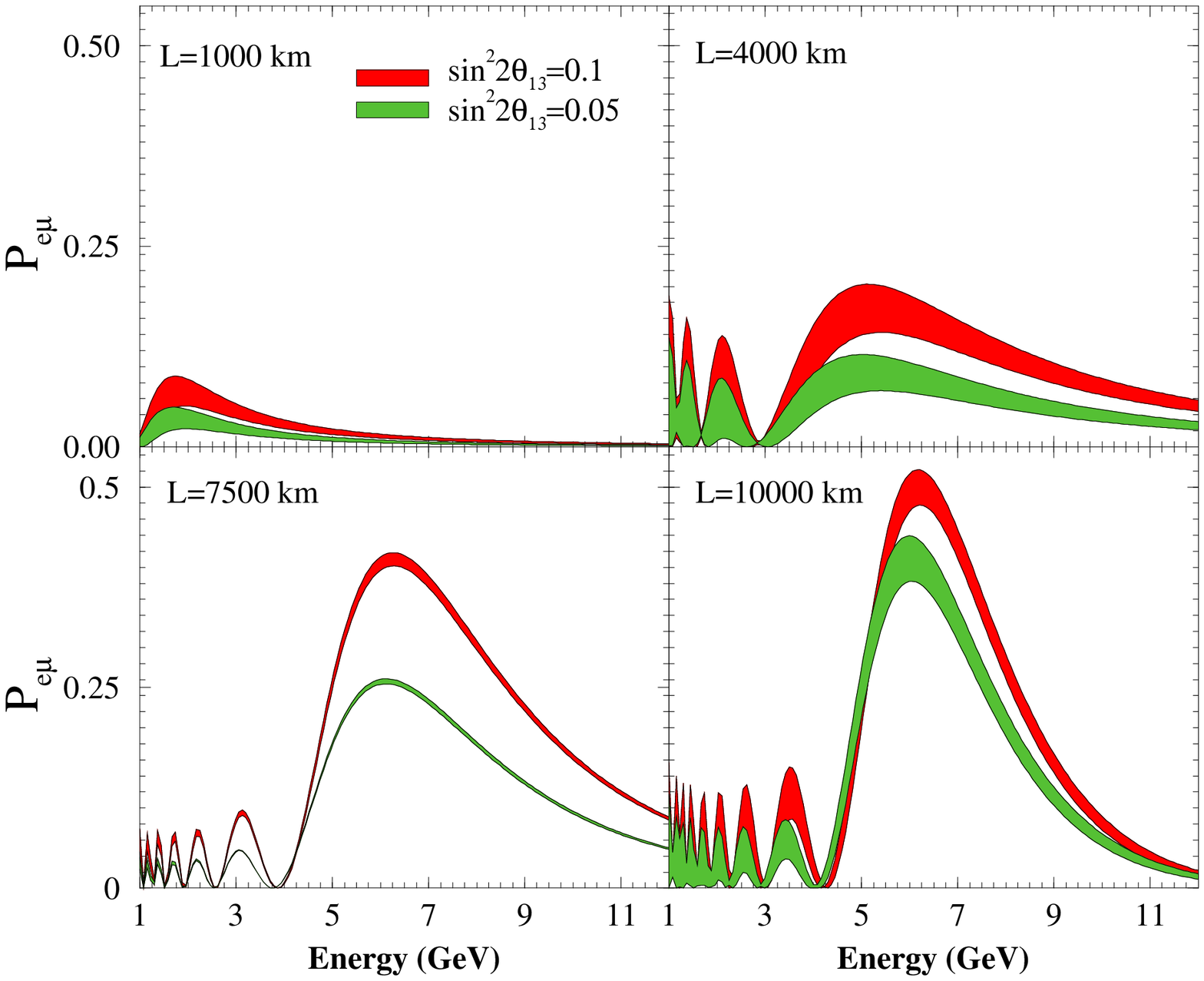}

\caption{\label{fig:prob}
Both the panels show the energy dependence of $P_{e\mu}$
for four baselines where the band reflects the effect of the
unknown $\delta$. Left panel clearly depicts the effect of $\delta$
in making distinction between normal (NH) $\&$ inverted (IH) hierarchy 
with $\stch=0.1$. Right panel reflects the difference in $P_{e\mu}$ 
for two different values of $\stch$ with normal hierarchy. All other
oscillation parameters are at their best-fit.}
\end{figure}

This large baseline requires traversal through denser regions of the earth.
Thus, for neutrinos (antineutrinos) with energies in the
range 3-8 GeV sizable matter effects are
induced if the mass hierarchy is normal (inverted). A unique
aspect of this set-up is the possibility of observing
near-resonant matter effects in the $\nue\rightarrow \numu$
channel. In fact, for this baseline, the average
earth matter density calculated using the PREM
profile is $\rho_{av}=4.17$ gm/cc, for which
the resonance energy is
$E_{res} \equiv |\ma| \cos 2\theta_{13} / 2\sqrt{2} G_F N_e
= 7.45$ GeV, taking $|\ma|=2.5\times 10^{-3}$ eV$^2$ and $\stch=0.1$.

Throughout this paper we show all our results assuming certain
true values for the oscillation parameters \cite{limits}
$|\Delta m^2_{31}| = 2.5 \times 10^{-3} \ {\rm eV}^2$,
$\sin^2 2 \theta_{23} = 1.0$,
$\Delta m^2_{21} = 8.0 \times 10^{-5} \ {\rm eV}^2$,
$\sin^2\theta_{12} = 0.31$ and $\delta_{CP} = 0$.
The exact neutrino transition probability using the PREM density 
profile is given in Fig. \ref{fig:prob}. As discussed above, for
$L=7500$ km, which is close to the magic baseline, the
effect of the CP phase is seen to be almost negligible
which allows a clean measurement of $\theta_{13}$ 
(see right panel of Fig. \ref{fig:prob}),
while for all other cases the
impact of $\delta_{CP}$ on $P_{e\mu}$ is
appreciable. As the baseline is increased, earth matter
density increases, enhancing the impact of matter effects.
The probability for normal hierarchy is hugely enhanced for
the neutrinos, while for the inverted hierarchy, matter effects
do not bring any significant change. This difference in
the probability can be used to determine the neutrino 
mass hierarchy (see left panel of Fig. \ref{fig:prob}).

\begin{figure}[!t]

\includegraphics[height=.24\textheight]{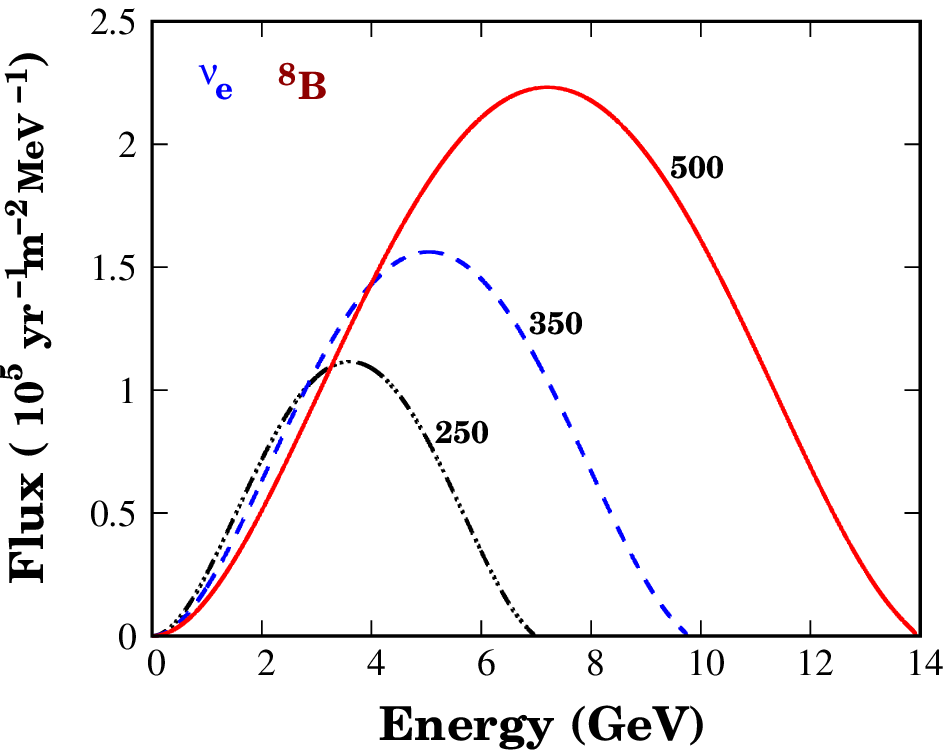}
\includegraphics[height=.24\textheight]{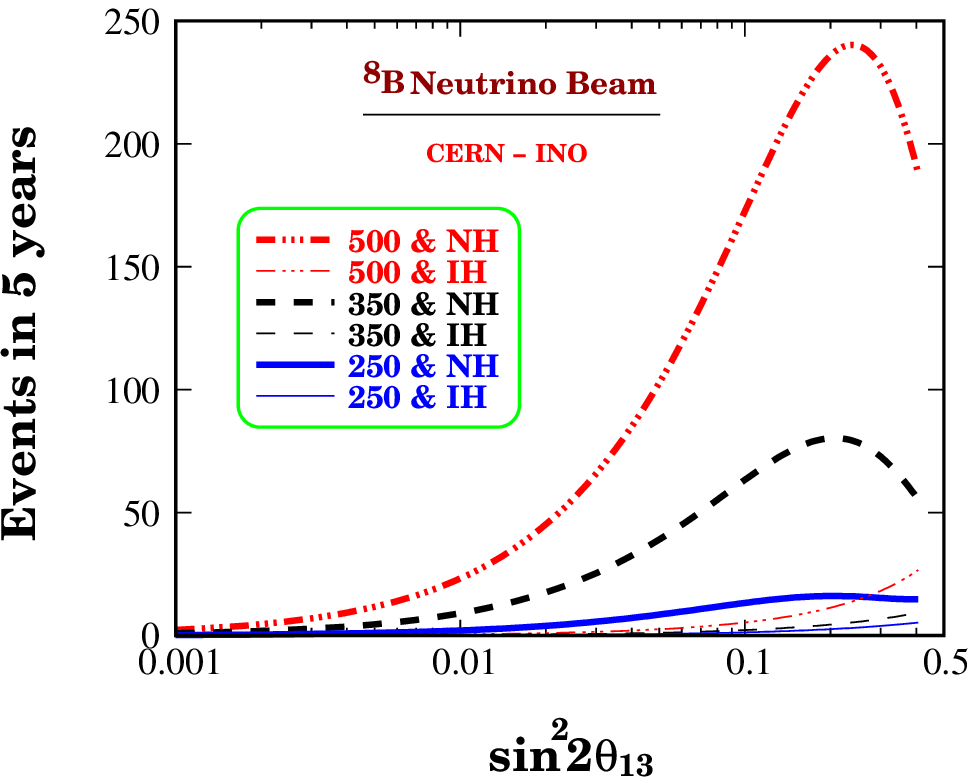}

\caption{\label{fig:flux_rate}
Left panel shows the boosted unoscillated spectrum of neutrinos 
from $^8B$ ion which will hit the INO detector, for three different 
benchmark values of $\gamma$. The expected number of $\mu^{-}$ events in
5 years running time with 60\% detection efficiency as a function of 
$\stch$ are presented in right panel. The value of $\gamma$ and
the hierarchy chosen corresponding to each curve is shown in the
figure legend.}
\end{figure}

Pure $\nu_e/\bar{\nu}_e$ beams can be produced from
completely ionized radioactive ions accelerated to
high energy, decaying through the beta process
in a storage ring, popularly known as
$\beta$-beams \cite{zucc, betapemu, abhijit, betaino, rathin, acgr}.
We consider $^{8}{B}$ ($^{8}{Li}$) \cite{rubbia}
ion as a possible source for a $\nu_e$ ($\bar{\nu}_e$)
$\beta$-beam. The end point energies of
$^{8}{B}$ and $^{8}{Li}$ are $\sim$ 13-14 MeV.
For the Lorentz boost factor $\gamma=250(500)$
the $^{8}{B}$ and $^{8}{Li}$ sources have peak energy
around $\sim 4(7)$ GeV. We can see from Fig. \ref{fig:flux_rate} 
(left panel) that with $\gamma$ = 500, the $\nu_e$ spectrum peaks 
nearby $E_{res}$. We assume that it is possible to get 
$2.9\times 10^{18}$ useful decays per year for $^8Li$ and 
$1.1\times 10^{18}$ for $^8B$ for all values of $\gamma$.
The study of $\beta$-beam flux at a near detector in the context 
of probing lepton number violating interactions is presented in \cite{subhendu}.


The proposed large magnetized iron calorimeter  at
the India-based Neutrino Observatory \cite{ino}
is planned to have a total mass of 50 kton at startup, which might be
later upgraded to 100 kton. The INO facility is
expected to come up at PUSHEP (lat. North 11.5$^\circ$,
long. East 76.6$^\circ$), situated close to Bangalore
in southern India. This constitutes a baseline of
7152 km from CERN. The detector will be made of magnetized iron slabs with
interleaved active detector elements.
For ICAL, glass resistive plate chambers have
been chosen as the active elements. In this
proposal the detector mass is almost entirely ($>$ 98\%) due to
its iron content.

According to the detector simulation performed by the INO
collaboration, the detector energy threshold for $\mu^\pm$ is
expected to be around $\sim 1.5$ GeV and charge identification
efficiency will be about 95\%.
For all the numerical results presented in this paper,
we calculate the exact three generation oscillation
probability using the realistic PREM \cite{prem} profile for
the earth matter density.

Large resonant matter effects in the neutrino channel for
normal hierarchy drives the number of expected events\footnote{We assume
a Gaussian resolution function with $\sigma=0.15E$.} 
to very large values, compared to what would be expected for
inverted hierarchy (see right panel of Fig. \ref{fig:flux_rate}). 
In case of anti-neutrinos, the expectations
are opposite with normal and inverted hierarchy.
Discussion on backgrounds and statistical analysis
for this set-up can be found in \cite{betaino}.

We define the {\bf{sensitivity to}} ${\bf{\theta_{13}}}$ as the minimum value of 
$\stch$ which this experiment would be able to distinguish from $\stcht=0$.
At $3\sigma$, the CERN-INO $\beta$-beam set-up can constrain
$\sin^22\theta_{13} < {\bf{1.1\times 10^{-3}(2.1\times 10^{-3})}}$ with
${\bf{60\%}}$ detection efficiency and 10(5) years data in the neutrino mode assuming
normal hierarchy as true with $\gamma=500$. If we can enhance the detection
efficiency upto ${\bf{80\%}}$ then the improved upper bound will be
$\sin^22\theta_{13} < {\bf{8.4\times 10^{-4}(1.6\times 10^{-3})}}$
with 10(5) years of neutrino run. These results are presented 
after marginalizing over $|\ma|$, $\sta$, and $\delta$.
We expect similar results with $\bar\nu_e$ $\beta$-beam
assuming inverted hierarchy to be true.

Next we focus on how {\bf{sensitive}} this set-up 
is {\bf{to}} ${\bf{sgn}}${\bf{$(\ma)$}}.
The minimum value of $\stcht$
for which one can rule out inverted hierarchy
at $3\sigma$ C.L. with 10(5) years of neutrino run assuming normal hierarchy
as true hierarchy with $\gamma=500$ and ${\bf{60\%}}$ detection efficiency
is ${\bf{8.7\times 10^{-3}(1.0\times 10^{-2})}}$.
For ${\bf{80\%}}$ detection efficiency, the improved numbers will be
${\bf{8.5\times 10^{-3}(9.8\times 10^{-3})}}$ with 10(5) years of neutrino run.
We marginalize over $|\ma|$, $\sta$, $\delta$
and as well as $\stch$ in calculating these numbers.
Similar performance can be expected with $\bar\nu_e$ $\beta$-beam
assuming inverted hierarchy to be true.

In summary, the current note discussed an experimental set-up with
a \bb source ($\nu_e$ or $\bar{\nu}_e$) at CERN and a
large magnetized iron detector at INO. The CERN-INO distance is 
very close to the ``magic" baseline which ensures a degeneracy-free 
measurement of the mixing angle $\theta_{13}$ and the neutrino mass ordering.
At this baseline we can get near-resonant matter effect 
for $E \approx 7.5$ ~GeV which is achievable with 
$^{8}{B}$ $\&$ $^{8}{Li}$ ions using $\gamma \sim 500$. 
The increase in the probability $P_{e\mu}$
due to near-resonant matter effects, compensates for the fall in
the \bb flux due to the very long baseline, so that one can
achieve sensitivity to $\theta_{13}$ and mass hierarchy which is
comparable, even better, than most of the other proposed experimental
set-ups.

\end{document}